\documentclass[amssymb,prl,twocolumn,showpacs]{revtex4}
\usepackage{epsfig}
\usepackage{dcolumn}
\usepackage{amsmath}
\begin{document}
\title{\bf Polarization immunity of magnetoresistivity response under microwave excitation}
\author{Jes\'us I\~narrea$^{1,2}$ and Gloria Platero$^2$}
 \affiliation {$^1$Escuela Polit\'ecnica
Superior,Universidad Carlos III,Leganes,Madrid,Spain\\
$^2$Instituto de Ciencia de Materiales, CSIC,
Cantoblanco,Madrid,28049,Spain.}
\date{\today}
\begin{abstract}
We analyze theoretically the dependence of the longitudinal
magneto-resistivity response of microwave irradiated
two-dimensional electron systems on the microwave polarization.
Both linear and circular polarizations are considered.
Recent experiments show that resistivity oscillations and zero
resistance states are unaffected by changing the polarization of
the microwave field. We propose a plausible explanation for the
experimentally observed magneto-resistivity polarization immunity.

\end{abstract}
\maketitle

Recently, magneto-transport experiments on two-dimensional electron
systems(2DES) irradiated with microwaves have shown two very
interesting features: Microwave Induced Resistivity Oscillations
(MIRO)\cite{zudov,studenikin} and Zero Resistance States
(ZRS)\cite{mani,zudov2}. This has motivated intense activity, both
experimental and theoretical. New and remarkable experimental
contributions are being published on a continual basis. Among them,
one can note the activated temperature dependence in the
magneto-resisitivity ($\rho_{xx}$)
response\cite{mani,zudov2,willett}, quenching of $\rho_{xx}$ at high
microwave (MW) intensities \cite{mani2,studenikin}, absolute
negative conductivity\cite{willett,mani2,studenikin,zudov3},
suppression of MIRO and ZRS by in-plane magnetic field
\cite{yang,mani3}, and the behavior of $\rho_{xx}$ under bichromatic
MW radiation coming from two monochromatic sources with different
frequencies\cite{zudov4}. Possibly the observation that MIRO and ZRS
are notably immune to the polarization of MW radiation in
Ref[\cite{smet}] is one of the most surprising results. In this
experiment the influence of the MW polarization on $\rho_{xx}$ in a
2DES was analyzed. Different MW polarizations were used, circular in
both senses (left and right) and also linear in x (current
direction) and y directions. The unexpected result of almost
complete immunity of $\rho_{xx}$ with the polarization was obtained.
All these new experimental results provide new and real challenges
for the theoretical models presented to date
\cite{girvin,dietel,lei,ryzhii,rivera,shi,andreev,ina,mirlin,kunold,auer}.

Some theoretical contributions have been presented that can explain
some of the new experimental outcomes. We can note proposals for
explaining the behavior of the MW driven $\rho_{xx}$ with
temperature and with high MW intensities \cite{ina2,lei2}, the
observed absolute negative conductivity\cite{ina3,ahn}, and the
$\rho_{xx}$ response to bichromatic MW radiation\cite{ina4,lei3}.
Regarding MW-polarization immunity, while some theoretical models
predict strikingly different dependences on the radiation
polarization\cite{andreev,mirlin,lei}, in others only linear
radiation was considered \cite{rivera,shi,ina}.

 In this letter we propose a theoretical
explanation for the experimental evidence which shows that
 $\rho_{xx}$ does not depend on the MW polarization. Our theoretical results are based
on the {\it driven Larmor orbits model}\cite{ina}. In a recently
presented work by the authors\cite{ina}, it was shown that in a 2DES
subjected to a moderate perpendicular magnetic field and MW
radiation, the Larmor orbit centers oscillate back and forth in the
$x$ direction with the same frequency as the MW field. A major and
non-trivial extension of this model \cite{ina} is presented here
which allows different polarizations for the MW field to be
considered, namely elliptical, circular or linear.

The results presented in this letter can be applied and
generalized to any physical situation consisting of a quantum
mechanical oscillator excited by any time-dependent force. We can
cite, for instance, nano-electromechanical systems (NEMS),
molecular electronics, surface acoustic waves (SAW) in Hall bars,
vibrational and rotational molecular spectra, etc. In the case of
a harmonic time-dependent force the physics obtained is even
richer, because new resonance situations arise.

Our system consist in a 2DES subjected to MW radiation, that can be
in different polarization states, a perpendicular magnetic field
($B$) (z-direction) and a DC electric field ($E_{dc}$)
(x-direction). In order to introduce the model we consider initially
left-circularly polarized MW, i.e., the electric field
$\overrightarrow{E}(t)$ of the MW radiation is:
\begin{equation}
\overrightarrow{E}(t)=E_{0}(\overrightarrow{i}\cos
wt+\overrightarrow{j}\sin wt)
\end{equation}
(a detailed mathematical analysis for other
polarizations will be presented elsewhere\cite{ina5}). $E_{0}$ and
$w$ are the amplitude and frequency of the MW field respectively.
The total hamiltonian $H$, working with the symmetric gauge for the
vector potential of $B$:
$(\overrightarrow{A_{B}}=-\frac{1}{2}\overrightarrow{r}\times
\overrightarrow{B})$, can be written as:
\begin{eqnarray}
H&=&\frac{P_{x}^{2}+P_{y}^{2}}{2m^{*}}+\frac{w_{c}}{2}L_{z}+\frac{1}{2}m^{*}\left[\frac{w_{c}}{2}\right]^{2}
\left[(x-X)^{2}+y^{2}\right]\nonumber \\
& &-\frac{e^{2}E_{dc}^{2}}{2m^{*}\left[\frac{w_{c}}{2}\right]^{2}}\nonumber-eE_{0}\cos wt (x-X)-eE_{0}x\sin wt \nonumber \\
& &-eE_{0}\cos wt X \nonumber\\
&=&H_{1}-eE_{0}\cos wt X
\end{eqnarray}
 $X=\frac{eE_{dc}}{m^{*}(w_{c}/2)^{2}}$ is the center of the orbit for
the electron cycloidal motion, $e$ is the electron charge, $E_{dc}$
is the DC electric field in the current direction, $w_{c}$ is the
cyclotron frequency and $L_{z}$ is the z-component of the electron
total angular momentum. $H_{1}$ can be solved exactly after lengthy
algebra\cite{ina5}, and using this result allows an $exact$ solution
for the electronic wave function of $H$ to be obtained:
\begin{eqnarray}
&&\Psi(x,y,t)=\phi_{N}\left[(x-X-a(t)),(y-b(t)),t\right]\nonumber  \\
&&\times  \exp \frac{i}{\hbar} \left[m^{*}\left(\frac{d
a(t)}{dt}x+\frac{d b(t)}{dt}y\right)+
\frac{m^{*}w_{c}(b(t)x-a(t)y)}{2}-\int_{0}^{t} {\it L} dt'\right]\nonumber  \\
&&\times\sum_{p=-\infty}^{\infty} J_{p}(A_{N}) e^{ipwt}
\end{eqnarray}
where $\phi_{N}$ are the analytical solutions for the Schr\"{o}dinger
equation with a two-dimensional (2D) parabolic confinement, known as
Fock-Darwin states\cite{fock}. The Fock-Darwin states converge to a
Landau level spectrum when $B$ is large or it is the only source of
confinement (present case). In polar coordinates
$\phi_{N}(r,\theta,t)$ can be expressed as:
\begin{equation}
\phi_{N}=\sqrt[]{\frac{n!}{2\pi
l_{B}^{2}2^{|m|}l_{B}^{2|m|}(n+|m|)!}}r^{|m|}e^{-im\theta}
L_{n}^{|m|}\left(\frac{r^{2}}{2l_{B}^{2}}\right)e^{-\left(\frac{r^{2}}{4l_{B}^{2}}\right)}
\end{equation}
where $n$ is the radial quantum number, $m$ is the angular momentum
quantum number, $L_{n}^{|m|}$ are the associated Laguerre
polynomials and $l_{B}$ is the effective magnetic length. For the
polar coordinates: \\
$r^{2}=[x-X-a(t)]^{2}+[y-b(t)]^{2}$ and\\
$re^{i\theta}=[x-X-a(t)]+i[y-b(t)]$. \\
$a(t)$ (for the x-coordinate) and $b(t)$ (for the y-coordinate)
are the solutions for a $classical$ driven 2D harmonic oscillator
(classical uniform circular motion). The expressions in the case
of a left polarized MW radiation are:
\begin{eqnarray}
a(t)&=&\frac{e E_{o}}{m^{*}\sqrt{w^{2}(w_{c}-w)^{2}+\gamma^{4}}}\cos
wt=A_{-}\cos wt \nonumber\\
b(t)&=&\frac{e E_{o}}{m^{*}\sqrt{w^{2}(w_{c}-w)^{2}+\gamma^{4}}}\sin
wt=A_{-}  \sin wt
\end{eqnarray}
\begin{figure}
\centering\epsfxsize=3.6in \epsfysize=5.2in
\epsffile{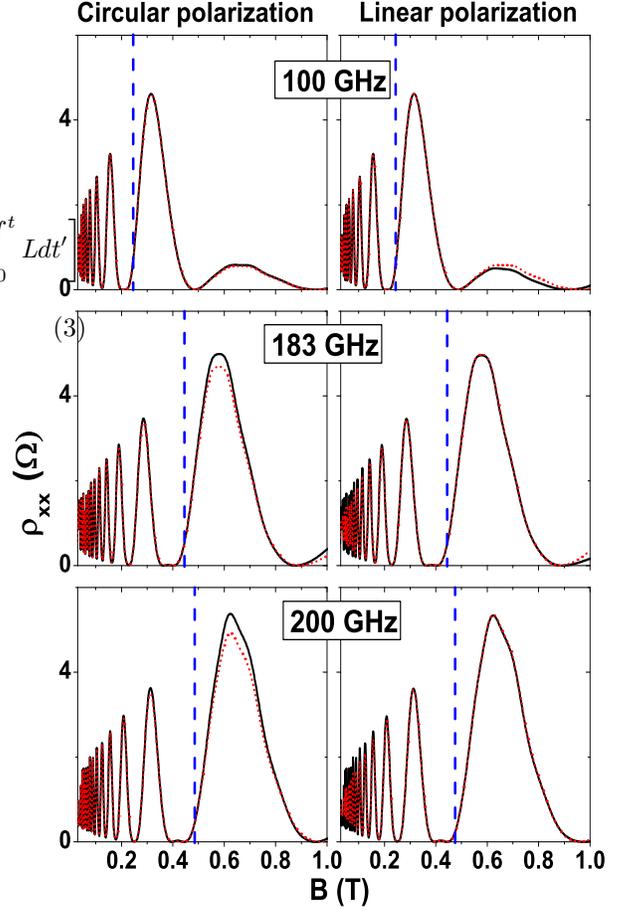} \caption{Calculated magneto-resistivity
$\rho_{xx}$ as a function of $B$ for circularly and linearly
polarized radiation for the experimental frequencies used, $100,
183$ and $200 GHz$. In the case of circular light (left panels)
the two senses of polarization have been considered: left, single
line (black color online) and right, dotted line (red color one
line) . For linear MW (right panels), we have considered the x
direction, single line (black color online) and y direction,
dotted line (red color online).
The $\rho_{xx}$ polarization immunity
can be observed clearly for the three frequencies,
especially for $B$ below the cyclotron resonance (see vertical
dashed line, blue color online). T=1K.}
\end{figure}
$\gamma$ is a material and sample-dependent damping factor which
dramatically affects the movement of the MW-driven electronic orbits, and
which has been introduced phenomenologically. Along with this
movement interactions occur between electrons and lattice
ions, yielding acoustic phonons and producing a damping effect in
the electronic motion. In Ref.\cite{ina2}, we developed a
microscopical model to calculate $\gamma$, estimating a numerical
value of $\gamma\simeq10^{12}s^{-1}$ for GaAs. ${L}$ is the
classical lagrangian:
\begin{eqnarray}
L&=&\frac{m}{2}\left[\left(\frac{d
a(t)}{dt}\right)^{2}+\left(\frac{d
b(t)}{dt}\right)^{2}\right]\nonumber\\
&&-m\frac{w_{c}}{2}\left[a(t)\frac{d b(t)}{dt}+b(t)\frac{d
a(t)}{dt}\right]
\end{eqnarray}
 and  $J_{p}$ are Bessel
functions , whose arguments, $A_{N}$, are given by (for left circular
MW):
\begin{equation}
 A_{N}=\frac{eE_{0}}{\hbar} X
\left(\frac{1}{w}+\frac{w+w_{c}}{\sqrt{(w_{c}^{2}-w^{2})^{2}+\gamma^{4}}}-
\frac{w_{c}(w+w_{c})}{2w\sqrt{(w_{c}^{2}-w^{2})^{2}+\gamma^{4}}}\right)
\end{equation}

Similar, but not identical, expressions for $\Psi(x,y,t)$ are obtained for
right and linear (x or y direction) polarized MW
radiation\cite{ina5}. The key differences among them are given by
the expressions we obtain for a(t) and b(t). Thus, for instance, for
right polarized MW light:
\begin{eqnarray}
a(t)=\frac{e E_{o}}{m^{*}\sqrt{w^{2}(w_{c}+w)^{2}+\gamma^{4}}}\cos
wt =A_{+}   \cos wt\nonumber\\
b(t)=\frac{-e E_{o}}{m^{*}\sqrt{w^{2}(w_{c}+w)^{2}+\gamma^{4}}}\sin
wt  = A_{+}   \sin  wt.
\end{eqnarray}

The first important result we obtain is that, apart from phase
factors, the wave function for $H$ is the same as a Fock-Darwin
state where the center of the electron orbits performs a circular
motion in the xy plane with frequency $w_{c}$, given by
$a^{2}(t)+b^{2}(t)=A_{\mp}^{2}$, ($"-"$ for the case of left
circular MW and $"+"$ for right circular light).

Electrons suffer scattering due to charged impurities that are
randomly distributed in the sample. If the scattering is weak, we
can apply time dependent first order perturbation theory.  To
proceed we calculate the impurity scattering transition rate $
W_{N,M}$ from an initial state $\Psi_{N}(x,y,t)$, to a final state
$\Psi_{M}(x,y,t)$\cite{ina,ridley}:
\begin{equation}
W_{N,M}=\lim_{\alpha\rightarrow 0} \frac{d}{d t} \left|
 \frac{1}{i \hbar} \int_{-\infty}^{t^{'}}<\Psi_{M}(x,t) |V_{s}|\Psi_{N}(x,t)>e^{\alpha t}d t\right|^{2}
\end{equation}
where $V_{s}$ is the scattering potential for charged
impurities\cite{ando}. After some algebra we obtain for this
transition rate $W_{N,M}$:
\begin{equation}
W_{N,M}=\alpha\frac{e^{5}n_{i}BS}{16\pi^{2}\epsilon^{2}\hbar^{2}}\left[\frac{\Gamma}{[\hbar
w_{c}(N-M)]^{2}+\Gamma^{2}}\right]
\end{equation}
$\Gamma$ is the Landau level broadening, $n_{i}$ is the impurity
density, $S$ the surface of the sample and $\epsilon$ the dielectric
constant. Since Coulomb interactions are not the primary focus
of this work, we will use the dimensionless parameter $\alpha$ to control
the strength of the electron-charged impurity interaction, rather
than considering a more sophisticated functional form. With
$\alpha=1$ representing the "bare" Coulomb interaction,
using a screened scattering potential corresponds
to $\alpha<1$\cite{pavel}.

The next step is to find the average effective distance advanced by
the electron in every scattering jump which is given by (see Ref.
[19] for a detailed explanation): $\Delta X^{MW}=\Delta X^{0}+
A_{\mp}\cos w\tau$, where $\Delta X^{0}$ is the effective distance
advanced when there is no MW field present and $1/\tau=W_{N,M}$
($\tau$ being the impurity scattering time). The magnitude $A_{\mp}$
is the amplitude of the orbit center motion in the x-direction
(a(t)):
\begin{equation}
A_{\mp}=\frac{e E_{o}}{m^{*}\sqrt{w^{2}(w_{c}\mp w)^{2}+\gamma^{4}}}
\end{equation}
 Considering that we have a static electric field
$E_{dc}$ in the current direction (x direction), we obtain an
average value (over all the scattering processes) for $\triangle
X^{MW}$ different from zero in that direction. Therefore the
electron possesses an average drift velocity $v_{N,M}$ in the x
direction. This drift velocity can be readily calculated by
introducing the term $\triangle X^{MW}$ into the expression of the
transition rate $W_{N,M}$, and finally the longitudinal (or
diagonal) conductivity $\sigma_{xx}$ can be
obtained\cite{ina,ridley}.
 To calculate $\rho_{xx}$ we use the relation
$\rho_{xx}=\frac{\sigma_{xx}}{\sigma_{xx}^{2}+\sigma_{xy}^{2}}$,
where $\sigma_{xy}\simeq\frac{n_{i}e}{B}$ and
$\sigma_{xx}\ll\sigma_{xy}$.

All our results have been based on parameters corresponding to
experiments by Smet et al. \cite{smet}. In Fig.1 we show $\rho_{xx}$
obtained using our model as a function of $B$ for circularly and
linearly polarized radiation and for the experimental frequencies.
In the case of circular light (left panels) the two senses of
polarization have been considered. For linear MW we have considered
the x and y direction (right panels). As in experiments, the
calculated $\rho_{xx}$ response is practically immune to the
polarization sense of circularly polarized MW radiation, specially
for $B$ below cyclotron resonance (see vertical dashed line). We
obtain similar results in terms of oscillations and ZRS for the
different polarizations considered. This behavior is observed for
all the different MW frequencies studied. When we introduced
linearly polarized radiation, $\rho_{xx}$ is almost the same if it
is linearly polarized in the x or y direction. According to our
model, the $\rho_{xx}$ response under MW excitation is governed by
the term $A_{\mp}\cos w\tau$: $\rho_{xx} \propto A_{\mp} \cos
w\tau$, where the amplitude $A_{\mp}$ has been defined above (see
eq. (11)). Therefore for left and right circularly polarized
radiation we would expect different results since the amplitudes are
different. However if the damping factor $\gamma$ is larger than the
MW frequency, $\gamma>w$, $\gamma$ would become the leading term in
the corresponding denominator of the amplitude $A_{\mp}$. In this
situation, $\gamma$ is able to quench the influence of the other
terms and similar values are obtained for the amplitude of the orbit
center for different polarizations. The same argument can be applied
to both linearly and circularly polarized MW radiation. For GaAs, a
value for $\gamma$ about $3\times10^{12}s^{-1}$\cite{ina2}, is
enough to obtain a similar $\rho_{xx}$ response irrespective of the
specific MW polarization.

In conclusion, we have theoretically studied the influence of the
microwave polarization on the magneto-resistivity response of
two-dimensional electron systems. Linear and circular polarization
have been considered with different senses. In agreement with
previous experimental results, we show that, under strong enough
damping, the $\rho_{xx}$ response is unaffected by changing the MW
polarization.

This work has been supported by the MCYT (Spain) under grant
MAT2005-06444 (JI and GP), by the Ram\'on y Cajal program (J.I.) by
the EU Human Potential Programme: HPRN-CT-2000-00144.

\end{document}